# A Family of Software Product Lines in Educational Technologies

Sridhar Chimalakonda[1]

*Indian Institute of Technology Tirupati*
*Renigunta Road, Settipalli Post, Tirupati, Andhra Pradesh, India - 517506*

Kesav V. Nori

*International Institute of Information Technology Hyderabad*
*Gachibowli, Hyderabad, Telangana, India - 500032*

**Abstract**

Rapid advances in education domain demand the design and customization of educational technologies for a large *scale* and *variety* of evolving requirements. Here, scale is the number of systems to be developed and variety stems from a diversified range of instructional designs such as varied goals, processes, content, teacher styles, learner styles and, also for eLearning Systems for 22 Indian Languages and variants.

In this paper, we present a family of software product lines as an approach to address this challenge of modeling a family of instructional designs as well as a family of *e*Learning Systems and demonstrate it for the case of adult literacy in India (287 million learners). We present a multi-level product line that connects product lines at multiple levels of granularity in education domain. We then detail two concrete product lines (`http://rice.iiit.ac.in`), one that generates instructional design editors and two, which generates a family of *e*Learning Systems based on flexible instructional designs. Finally, we demonstrate our approach by generating *e*Learning Systems for Hindi and Telugu languages (both web and android versions), which led to significant cost savings of 29 person months for 9 *e*Learning Systems.

## 1. Motivation

The role of technology in education has undergone a massive transformation in the 21st century promising to facilitate anywhere, anytime learning to everyone [2][3]. There is also a dramatic rise in the use and design of a variety of technologies in education in the last decade or so [4]. A plethora of educational technologies such as computer assisted instruction [5], web based learning [6], game based learning [7], learning management systems [8], computer-supported collaborative learning [9], virtual learning environments [10] have emerged for a wide range of environments and contexts across the globe. Even though these technologies vary on several dimensions, *software* is a central theme of many of these technologies. In addition, there is also a need to constantly improve these technologies catering to emerging trends such as personalized learning [11], gesture based learning [12], augmented reality [13], gamification [14] and Massive Open Online Courses (MOOCs) [15]. This further increases the complexity during the design of educational technologies and makes it an incredibly hard challenge to *customize* and *adapt* these technologies as per the emerging trends and requirements.

On the other hand, there is also severe criticism on several dimensions such as huge upfront costs of technologies, difficulties in using them on the field, lack of evidence to show positive impact on quality of education [16]. One major challenge was an ever increasing effort required to develop and maintain a large number of educational technologies, which often ends up as an overburden on the teachers [16].

In essence, this scenario poses grand challenges for engineering and computing such as (i) *Advance personalized learning*[2] and (ii) *Provide a teacher for every student*[3]. From a technological perspective, the challenge is to facilitate *design and customization of educational technologies for scale and variety*, where *scale* is the number of systems to be developed and *variety* stems from variations in all aspects of teaching and learning.

---

✩This paper contains content from first author's PhD thesis [1]
[1]*ch@iittp.ac.in*, This work was done when the author was at IIIT Hyderabad, India for his doctoral thesis

[2]Grand Challenge 14 from *Grand Challenges for Engineering, National Academy of Engineering of the National Sciences, https://goo.gl/U8FpKv*
[3]Grand Challenge 3 from *Grand Research Challenges in Information Systems, Computing Research Association, https://goo.gl/ciDVa1*

This challenge is further exacerbated in Indian context owing to the need to design and customize educational technologies for 22 Indian Languages and variants for a wide range of learners, teachers and in varied contexts. This was also echoed in a steering committee meeting of planning commission[4] which suggested that any technology-based solutions should be (i) based on strong pedagogical basis (ii) available for all Indian languages and (iii) flexible to cover varied teaching styles, different types of learners and varied Indian contexts. Most importantly, the primary concern was in terms of the cost and effort required for creating and maintaining the technologies for the scale and variety inherent for education in India.

To this end, *the core contribution of this paper is a software product line approach to the design and customization of educational technologies for scale and variety in the domain of education and specifically adult literacy in India.*

The paper is organized as follows. We explain the core concepts and definitions in next section. In Section§3, we present the challenge of scale and variety in the context of adult literacy in India and outline the technological requirements. The need for software reuse and software product line approach is motivated through literature in Section§4 followed by a brief overview of the proposed approach in Section§5. We introduce a family of different product lines in education domain in Section§6. A multi-level software product line covering the scope of product lines in this paper is presented in Section§7. We present a software product line for a family of instructional designs in Section§8 along with feature models, feature attributes, feature configurations in subsequent sections. We then discuss a reference architecture in Section§8.4 that is used as the base for designing prototype platforms for instructional design and *e*Learning Systems. In Section§9, we give two concrete examples of *e*Learning Systems for *Hindi* and *Telugu* language generated from our prototype platforms. The experimental results and cost savings of using our approach are given in Section§10. We finally end the paper with conclusions and future work of applying software product lines for personalized learning.

## 2. Key Concepts & Definitions

Berger defines instructional design as a *"systematic development of instructional specifications using learning and instructional theory to ensure the quality of instruction"* [17]. In this paper, we consider *instructional design as an underlying structure consisting of different aspects of instruction such as goals, process, content aimed at (i) providing a base for quality of instruction (ii) facilitating design of educational technologies*

*Educational Technologies* - Educational technologies is a broad concept [18] sometimes used interchangeably with terms such as learning technologies and instructional technologies. One common definition from Association for Educational Communications and Technology (AECT) is *"Educational technology is the study and ethical practice of facilitating learning and improving performance by creating, using, and managing appropriate technological processes and resources"* [19]. According to Council for Educational Technology, UK, *"Educational technology can be considered as the systematic development, application and evaluation of systems, techniques and aids to improve the process of human learning."*

We consider *"educational technologies as a set of processes, techniques, methods and tools that facilitate systematic development of eLearning Systems based on well-established instructional designs."*

*eLearning Systems* - We consider *e*Learning Systems as a sub-class of educational technologies that are designed for improving learning and teaching in a particular context. Specifically, we consider *e*Learning Systems[5] as simple multimedia systems that use audio and visual aspects to teach reading, writing and basic arithmetic corresponding to physical instructional material in the context of adult literacy in India [20]. In essence, *educational technologies* facilitate systematic design of *eLearning Systems*.

## 3. The Scale & Variety Challenge - A Case Study

*How to facilitate design and customization of eLearning Systems to teach 287 million adult illiterates in India spread across 22 Indian Languages, who are beyond the age of schooling, earning their livelihood, who speak their native language, but cannot read or write?*

The world has undergone a rapid transformation into digital age with over an estimated 7 billion mobile users and around 2.4 billion Internet users worldwide [21]. However, the same world has an estimated 775 million young people and adults who are unable to read or write even in the digital era [22]. Surprisingly, India itself has around 37% of them, who are beyond the age of schooling, speak their language, but cannot read or write and spread across 22 Indian Languages [22].

In addition, according to reports from Government of India, the present average of adult illiterates taught by instructors is around $10^6$, whereas even assuming 200 adult illiterates per year for 5 years would still need a dedicated force of 287,000 instructors. The National Literacy Mission (NLM) of Government of India (GoI) has been striving to address this challenge since 1988 and has created a uniform methodology for teaching adult illiterates across India [23]. In the literature, there were several efforts of using technologies such as radio, television and even mobiles to reach out to adult illiterates in India [24] [25].

---

[4] *August 2011, Planning Commission* is headed by the Prime Minster of India and is later replaced by NITI AYOG, a think-tank for guiding the Government of India

[5] We also use the term *i*Primers for *e*Learning Systems in the context of adult literacy in India

[6] *August 2016*, Personal Communication, State Resource Center Director, Telangana, India under the programme of "Each one, teach ten"

A technology initiative by Tata Consultancy Services (TCS), an Indian Software Consultancy Services firm, as part of their Corporate Social Responsibility program consists of 9 *e*Learning Systems for 9 Indian Languages and has made around 120,000 people literate [26]. While these experiments have yielded significant productivity increase over Government of India efforts, with decreasing dropouts and increasing pass rates [26], the instructional design was constant and the *e*Learning Systems are *monolithic* in nature making their customization an open challenge. We have applied the ideas of software reuse and reduced the effort from 5 to 6 person years spread over 2 calendar years to 5 to 6 person months in 6 calendar months for developing an *e*Learning System [20]. But this approach is for automating a family of *e*Learning Systems based on a *fixed* instructional design whereas the dire necessity is to design *e*Learning Systems for flexible instructional designs and further customize them for 22 Indian Languages and variants. This presents the following key requirements and challenges, setting the context for this paper:

- Facilitate design of *e*Learning Systems for 22 Indian Languages and dialects *(scale)*.

- Facilitate the design of these *e*Learning Systems for flexible instructional designs (*varying* goals, processes and content) catering to the *varying* needs of 22 Indian Languages and variants.*(variety)*

- Facilitate the development of *instructional design editors* for creation of customizable instructional designs.

- Facilitate quality of instruction by basing the design of these *e*Learning Systems on instructional design.

Even though these challenges are specific to adult literacy, design of *e*Learning Systems for other forms of education such as schooling, skills, engineering, and customizing them for varied contexts and delivering them in multiple languages makes it a grand challenge. Table 1 shows an example possibility of designing *e*Learning Systems for six subjects from K1 to K12 with each of them having varied goals/process/content to be delivered in 22 Indian Languages and variants. The problem in these cases is of *scale* and *variety* during the design of these *e*Learning Systems for varied instructional designs.

In the next section, we present an overview of literature for design of educational technologies from a software engineering perspective.

## 4. Related Work

Explicit modeling of instructional design and its variants is a fundamental aspect to facilitate *scale* and *variety* inherent in the problem domain. In the context of this paper, we consider instructional design as an underlying structure that encompasses principles of instruction to facilitate design of educational technologies. There has been extensive research on modeling instructional design for the last several years resulting in a plethora of educational modeling languages (EMLs) [27] [28] [29] such as poEML [30], PALO [31], Web COLLAGE [32] as a way to model and reuse aspects of instructional design. Sampson et al. presented an open access hierarchical framework for integrating open educational resources at different levels of granularity [33]. IMS-LD emerged as a standard for learning design [34] and then focus shifted to tools such as LAMS [35] and LDSE [36] that aim to support teachers. A vision paper aimed to create an approach that integrates most of these tools towards an integrated learning design environment [37]. Despite this rapid progress, many researchers have pointed to several shortcomings of modeling and reusing instructional design such as complexity of authoring, lack of adequate tool support, interoperability and inability to support teachers [38]. In addition, several researchers have used ontologies as a means to represent different aspects of instructional design [31] and learning design using IMS LD [39]. LOCO [40] was presented as an ontology to bridge the gap between learning objects and learning designs through context. However, these ontologies and tools based on them are tightly coupled with each other and do not support for modeling instructional design variants making it difficult for design of *e*Learning Systems for *scale* and *variety*.

Table 1: Scope of Educational Technologies - An Example

| Class | 1st, 2nd, 3rd Language | Maths | Science | Social |
|---|---|---|---|---|
| K1 | $l_1...l_{22}$/Varying goals/process/content | $l_1...l_{22}$ | $l_1...l_{22}$ | $l_1...l_{22}$ |
| K2 | $l_1...l_{22}$/Varying goals/process/content | $l_1...l_{22}$ | $l_1...l_{22}$ | $l_1...l_{22}$ |
| K3 | ... | ... | ... | ... |
| K12 | $l_1...l_{22}$/Varying goals/process/content | $l_1...l_{22}$ | $l_1...l_{22}$ | $l_1...l_{22}$ |

Development of software components for the domain of education started way back in 1999 [41]. But the use of software engineering approaches in educational technologies has garnered significant attention with the advent of reuse of learning objects [42]. Designing reusable learning objects was extensively studied by several researchers [42][43][44] [45] [46][47] [48] in educational technologies but most of these efforts have not been very fruitful due to lack of emphasis on critical aspects of instructional design [49] [50] [51] [52] [53]. Design principles from software engineering were borrowed to facilitate reuse of learning objects [44]. However, this emphasis itself has led to severe criticism on software engineering being misused in the context of learning objects from a learning perspective [51]. Researchers have used model driven development to facilitate reuse of learning objects [54]. Dodero et al. further proposed a model-driven approach to learning design, a domain-specific language and a tool based on this approach to facilitate modeling of learning designs [55]. A model-driven development approach for learning design using the LPCEL Editor was proposed in [56]. But despite the advantages of these generative approaches, it was noted that the complexity of authoring process increases because of model development required from domain experts [57]. The term educational software engineering was coined in [58] but the focus has been on games for software engineering education.

One area of work that is directly relevant to this paper is called as Software Product Lines (SPLs) that facilitates systematic reuse across a family of systems [59]. Over the last two decades, there has been an extensive research on SPL as evidenced through a series of focused conferences such as Software Product Lines Conference (SPLC), workshops like Variability Modelling of Software-intensive Systems (VaMoS), Product Line Approaches in Software Engineering (PLEASE). An analysis of the literature on SPL reveals that are two major terminologies to discuss the idea of developing a family of software-intensive systems. Software Engineering Institute (SEI) has steered the research and development on software product lines (sometimes called as software product family) and has published several technical reports and case studies [59]. On the other hand, several researchers and organizations also used the term "software product line engineering" for their work in the area of SPL [60]. Several organizations, universities and research institutes performed collaborative research on SPL, which supported the systematic building of a community of software product line engineering research and practice. Some of those projects include ARES (1995-1998), PRAISE (1998-2001), ESAPS (1999-2001), CAFÉ (2001-2003), and FAMILIES (2003-2005). In 2014, Metzger and Pohl have done an extensive study of 600 articles published in the area of SPL and noted that there has been impressive quantitative and qualitative progress in the field with key challenges for industrial adoption [61]. Krueger has suggested three ways of adopting SPLs [62] (i) *proactive*, in which the entire product line is planned and developed from scratch (ii) *extractive*, that focuses on analyzing a set of existing products and moving towards an SPL (iii) *reactive*, that starts with one product and extends into an SPL. Depending on the product line strategy an organization can choose the appropriate product line adoption approach. There are several approaches for development and analysis of SPLs in the literature [63][60][64][65][66]. One related approach proposed for the domain of flight control focuses on using architecture and design patterns for SPL [67] but largely confines to modeling variability. Researchers have also used ontologies for modeling and configuring variability during SPL [68][69] [70]. A software product line is developed based on ontologies for developing knowledge-driven semantic web applications [71] to facilitate interoperability between semantic services and intelligent agents. A more detailed account of research in SPL can be found in [61] and a recent bibliographic analysis of research in SPL over 20 years is provided in [72].

On the other hand, many of the SPL approaches have been applied in the last couple of decades in practice across several domains with successful results [73]. However, there is sparse research on applying SPL in the domain of technology enhanced learning [74]. Pankratius has proposed PLANT as a product line based approach for creation and maintenance of digital information products [75]. In our prior work, we proposed TALES as an approach for automating the development of *e*Learning Systems [20]. A software product line methodology for development of e-learning system for a six sigma course was proposed in [76]. A domain engineering activity for interactive learning modules is proposed in [77]. [78]. A software product line for m-learning focusing on programming is discussed in [79]. However, none of these approaches consider instructional design domain as the basis and do not focus on scale and variety inherent in the problem domain of education. After a critical analysis of literature, we find that SPL is largely undermined in technology enhancing learning community despite their significant potential and hence motivating our approach.

*To summarize, existing approaches in the literature focus on either modeling instructional design from learning perspective or on software reuse and not both presenting a strong motivation and need for our approach.*

## 5. High-Level Overview of Proposed Approach

Design of educational technologies for *scale* and *variety* while maintaining quality is a major challenge requiring research from several disciplines such as learning methodologies, educational technologies, software engineering and human-computer integration. For the last several years, we have been working on creating several technological aids to support education in India with our research spanning across educational technologies [80][81], software engineering [82] and human computer interaction (HCI) [83]. We have briefly summarized some of these different perspectives in [84]. However, the focus of this paper is on applying software engineering approaches and principles to accelerate the design of educational technologies for *scale* and *variety* based on well-established learning methodologies and demonstrate it in the case of adult literacy. To this end, we rely on the following inputs from domain:

- An educational philosophy that provides a strong basis for learning and teaching.

- Instructional material devised by domain experts based on the above methodology.

- Field tested *e*Learning Systems based on this instructional material.

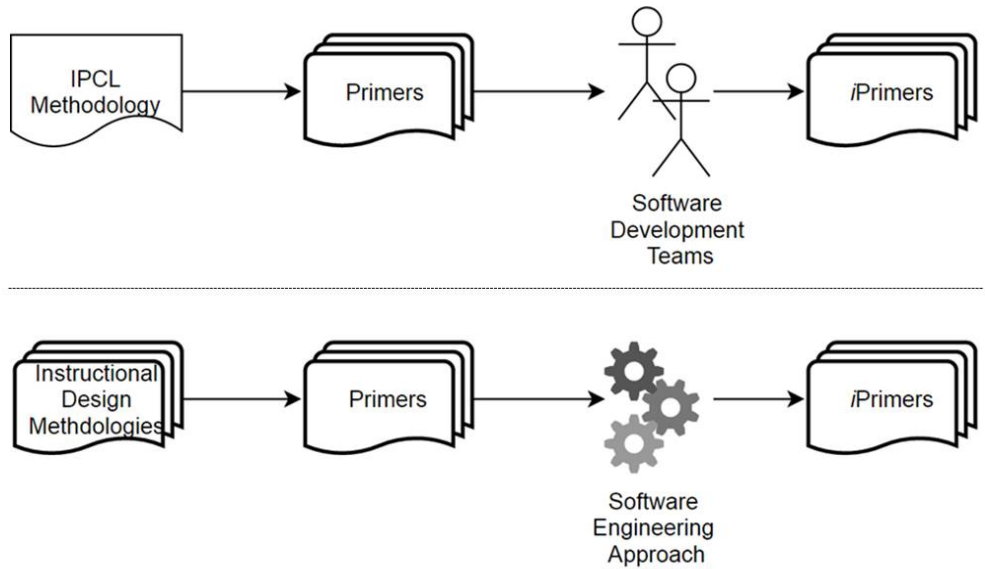

Figure 1: Existing and proposed approach for design of educational technologies for *scale* and *variety*

Figure §5 shows a simple schematic of the existing approach (top) and proposed approach (bottom) for design of *i*Primers for adult literacy in India. In the existing approach, individual software development teams develop *i*Primers for every primer and all these primers are based on a single instructional design methodology i.e., IPCL in the case of adult literacy in India. The core idea of the proposed approach is to systematically model different aspects of instructional design using *patterns* [85], concretely represent them using *ontologies* [86] and then apply a *software product lines* approach for semi-automatically generating *e*Learning Systems for varied instructional designs and multiple languages. The key difference is that the proposed approach can handle the *scale* and *variety* for flexible instructional designs instead of re-developing *e*Learning Systems for every new case and every change in the inputs and eventually allowing flexible modeling of instructional designs and creation of customizable *i*Primers.

In the next section, we motivate the need for software product lines in the context of educational technologies.

## 6. Families in Educational Technologies

The predominant way of developing software today is either a generalized product such as *Office*, *Gmail* or *Payroll* for a large number of customers or specific software that is designed for individual customers as in service based organizations. Consider the scenario of software for all banks? Is it same for all banks? Is it different for all banks? Similarly, if we consider software for all accounting systems? Is it same or different? The idea of looking at software systems as a family rather than completely different individual systems offers several advantages [87]. A family of systems can be at different levels of granularity with multiple sub-families within families and also a hierarchy of families. In this section, we discuss several families that are of interest to us in this paper. A *family* of systems can be defined as *a set of systems that share more common properties with other members in the set than differences providing unique advantages to address the common and varying needs of specific markets* [87].

### 6.1. Instructional Design as families

In this paper, we consider *Instructional Design* as a fundamental tenet that forms the basis for design of educational technologies. How is instructional design developed? Is it developed from scratch? or reused from existing resources? Are instructional designs common across subjects? learners? teachers? or universities? How many instructional design theories are present in the literature? What is common across them? Can instructional designs be considered as a family? Charles Reigeluth has extensively studied instructional design theories and models and documented them in three voluminous books [88][89][90]. His vision was to build a common knowledge base and a common language about instruction [88]. We can classify instructional design in the form of several families and at different levels of granularity as shown in Figure §2. For example, we can consider the three fundamental ways of *cognitivism*, *behaviourism* and *constructivism* or we can group them based on models such as *Gagne's model* [91], *Dick and Carey's model* [92], or the generic *ADDIE process* that is followed in most of the instructional designs. Each of these models can be grouped as a family based on subjects like STEM or K-12 or can also be formed as a family primarily based on learning styles such as *Visual, Kinesthetic, Auditory*. Here [A] and [B] in Figure §2 are at one level of granularity whereas [C] and [D] are at the next level of granularity. This can be further refined into a hierarchy of families till the lowest level of granularity. The classification primarily depends on the goals of the specific target organization or stakeholders. For example, if a university has professors who are keen on using *"learning by doing"* approach, then a family of instructional designs can be modeled with *"learning by doing"* as the base approach and adapting it for different kinds of learners and subjects. On the other hand, if a university policy mandates accreditation with a national body, then all instructional design should use principles of accreditation as base and then adapt them for specific needs. It is mandatory that all members belonging to a family have certain common properties and vary on some aspects making them fit for the specific purposes.

The core idea is not to look at every instructional design as a unique case but as a family of similar but distinct instructional designs, to leverage the common properties of the family and facilitate flexible instructional designs. Merrill has distilled a large number of instructional design models and came up with five fundamental principles that are common across many instructional design models [93] and there can be several variants based on these principles giving a family of instructional designs. We use the pattern categories identified in [85] as base for modeling instructional designs and variants. However, considering the generic scope of all instructional designs in the literature is out of scope of this paper.

### 6.2. Adult Literacy as families

Should there be a universal technological solution for teaching all 287 million adult learners across India? or Should there be unique solution for every learner? This leads to a trade-off and need for a balanced solution between *one-size-fits-all* and *unique-for-every-dimension* solutions. NLMA, the highest authority for adult literacy in India has devised a uniform methodology called IPCL as the base for creating instructional designs for all languages across India [23]. The handbook of IPCL provides guidelines for customizing different aspects of instruction based on several dimensions [23]. The key goals for adult literacy primers are to create immense interest in the learners and provide functional knowledge that can add value to learners' daily life [94]. These common goals have to be customized for learners spread across India based on socio-cultural and local contexts.

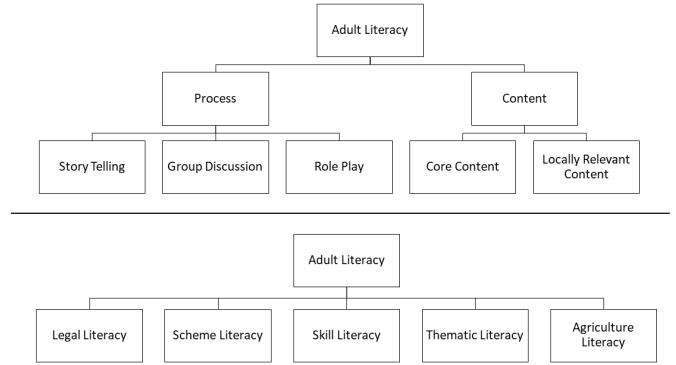

Figure 3: Few sample families of adult literacy domain

Based on [94] and [23], we show a portion of family for organization of adult literacy domain in Figure §3. Two common components of this family are *Process* and *Content*, where each one of them will have core aspects that are present in every family member and can also impose some constraints. The process followed for adult literacy in India is generally driven by *eclectic method* that starts teaching from known to unknown with gradual progress in learning. Now, any instructional design for adult literacy domain in India that comes under this family must follow *eclectic method* unlike *synthetic method* that teaches from alphabets to words. The process itself can have many number of activities such as *StoryTelling, GroupDiscussion, RolePlay* that can be customized based on specific needs. This family also says that *Content* should be present and can be further divided into *CoreContent* and *LocallyRelevantContent*. *CoreContent* mandates topics such as national integration, secularism, democracy, scientific temper, communal harmony, women's equality, population education and development, etc. whereas *LocallyRelevantContent* will be customized by specific states and stakeholders based on learner's livelihood, their socio-cultural realities, special issue-based and thematic aspects such as gender parity, health and hygiene, agricultural, animal husbandry, self-help groups, local self-government, livelihood programmes, etc. Then same family can be classified based on the primary *topic/knowledge* that can be used to teach knowledge that will be useful for learners in their daily life. *Legal Literacy* can focus on teaching learners laws pertaining to them, how to seek help from law whereas *Scheme Literacy* can provide knowledge of Government schemes. *Skill Literacy* can help them in gaining knowledge pertaining to a particular skill like tailoring, plumbing and so on. *Thematic Literacy* is a generic way to accommodate themes that can be local to the specific audience and *Agriculture Literacy* can provide farming knowledge. Each of these families can mandate that any adult literacy instructional design based on the corresponding family should be within the scope of the defined goals and constraints. These common themes are combined with specific needs of the particular segment of learners to deliver a specific and customized instructional design. These commonalities and variabilities among family members are generally modeled using feature diagrams in SPL [95][96].

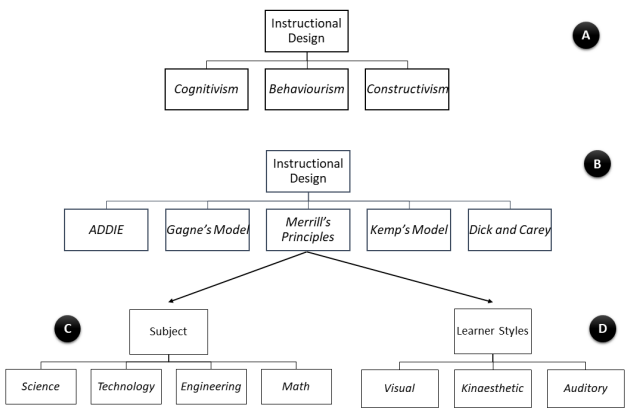

Figure 2: Few sample instructional design families

*6.3. eLearning Systems as families*

Eventhough we do not focus on modeling variants of user interface in this paper, it can be a critical source of variability for *e*Learning Systems. There exists certain ways to model user interfaces [97] as a hierarchy of Presentation Units that allow navigation between them and each of the units further contain UI elements such as buttons, textboxes and so on. These UI elements have properties such as *name*, *data* and so on. Each of these elements also have properties to describe their visual appearance such as *color*, *font* and so on. The user interface can be modeled for different platforms like *Desktop*, *Web*, *Mobile* or the user interface can also be modeled using *structural* and *behavioral* elements. These elements can be organized at a higher level of abstraction in different ways giving several user interface variants. For example, there can be several *View*s of Model-View-Controller pattern corresponding to different variants of user interfaces. In the case of adult literacy *e*Learning Systems, we are interested to model the user interface elements primarily with three resources *text*, *image*, *audio*.

In the next section, we present a multi-level software product line that connects multiple software product lines in this paper.

## 7. A Family of Software Product Lines

Figure §4[A] succinctly summarizes different levels of product lines that we considered in this paper. We reiterate our notion of instructional design as a set of *goals*, *process*, *content*, context, evaluation, environment and so on towards facilitating learning. Figure §4 [A] is a meta-level product line that deals with creating specific instructional designs from a chosen base instructional design. Here, there could be several sub-product lines focusing on a particular instructional design. For example, an instructional design like *learning by doing* [LBD] might be chosen as the base for all instructional designs in a particular university. Then, the derivations of LBD customized as per specific requirements of the courses in the university form a product family. Here, the input is a specification or schema of an instructional design and can consist of all features [including pre-requisites, activities, assessment and so on]. All product family members might not require all the features of LBD and hence only a subset of this instructional design specification is required for specific instructional design requirements. The scope of this meta product line is to create custom instructional design specifications based on a given instructional design specification. Similarly, there can be a number of sub-product families within this product line pertaining to a type of instructional design inquiry-based learning, IPCL for adult literacy and so on.

How to create instances of the custom instructional design specifications? Figure §4[B] shows a product line at the next level whose product family members are custom instructional design editors that take an instructional design schema[7]. We designed a prototype to generate these custom editors based on the specific instructional design specifications (instances). Each of these editors can be used to generate the concrete instructional designs with data. Even though motivated by adult literacy, these two product lines are in the context of generic instructional design. To co-relate with literature from educational technologies, these editors are similar in principle to learning design editors such as *ReLoad* and *ReCourse Editor* [98], *ASK-LDT Editor* [99], *LAMS* [35], *Learning Designer* [100], *COLLAGE* [101], *Web-COLLAGE* [32], *ILDE* [102] [103] and so on, where each of these editors are single system development initiatives as part of EU funded projects unlike the proposed product line approach.

Figure §4[C] shows the next level of product line that is specific to a custom instructional design specification, in this case one based on IPCL and adult literacy instructional design. We designed a prototype that takes a specific instance of adult literacy instructional design and generates *e*Learning Systems, which are the product family members for this product line.

Over the last decade or so, SPL community has witnessed a voluminous number of tools from academia as well as industry to support the entire software product line development life cycle [65][104]. We have primarily used two tool suites for modeling features in our SPL (i) *FeatureIDE* is developed on top of *Eclipse* and is quite useful as it supports multiple feature modeling techniques and also for generating code in several programming languages [105]. (ii) feature modeling plugin from University of Waterloo is a dated solution specifically useful for cardinal features and feature cloning and feature attributes [106]. For example, a *fact* in *ContentPattern* as a feature should be cloned for various instances. In their research, the same group has produced a minimalistic modeling language called *Clafer* and a set of tools as part of SPL platform [107]. The primary goal of *Clafer* is to address long standing concerns (merging feature and class models, mapping features to component configurations) in feature mod-eling by integrating feature modeling and meta modeling with rich semantics [107]. However, we realized that owing to the specific requirements from educational technologies domain, there is a strong need to extend the idea of feature attributes such that data pertaining to aspects in instructional designs can be annotated with feature models. For demonstration purposes, we have manually annotated features with concrete data for further processing by tools.

---

[7]A detailed listing of the instructional deign specification is available at https://git.io/vdxJG

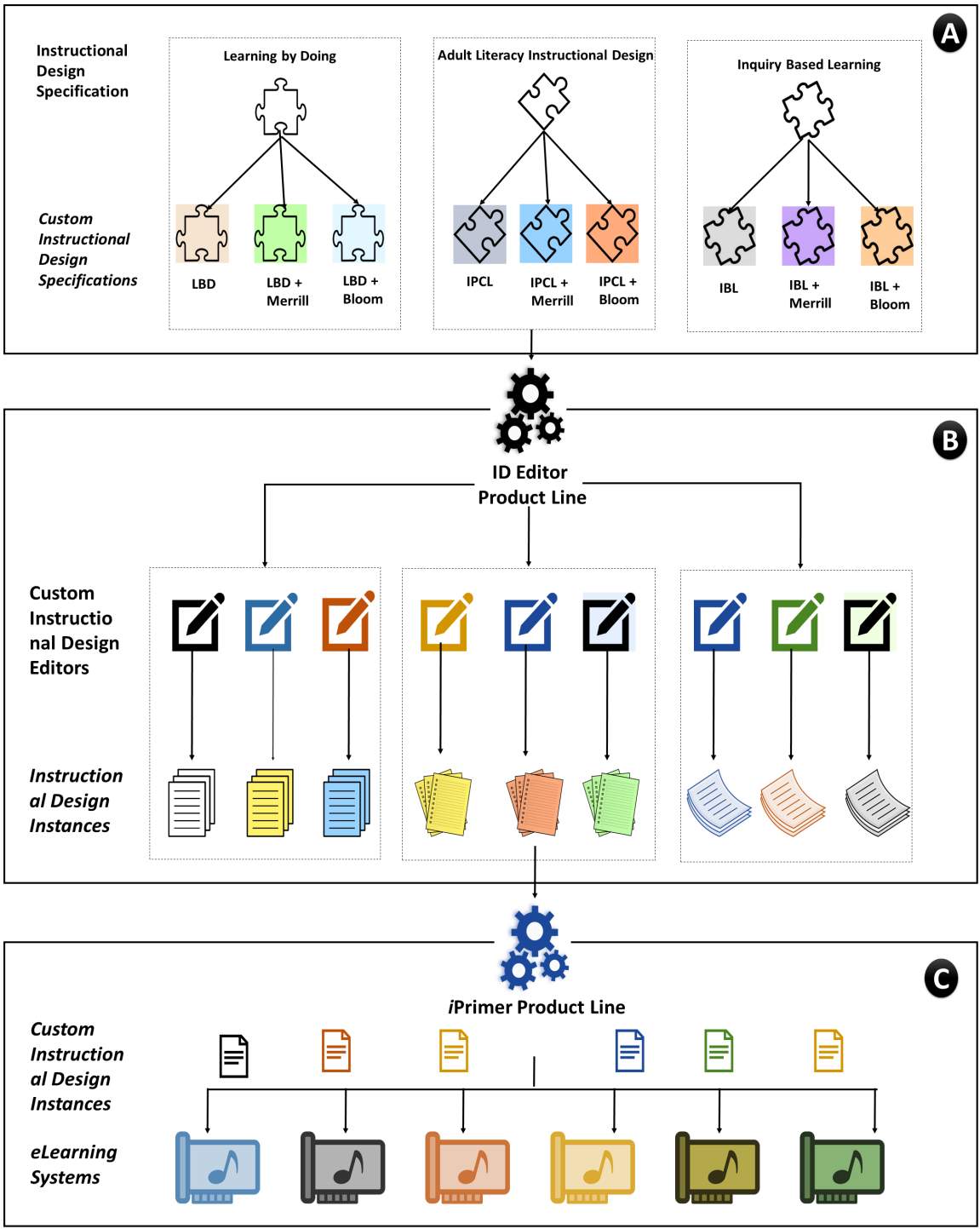

Figure 4: Multi-level software product lines

In the next sections, we succinctly describe the two product lines for instructional design and *e*Learning Systems.

## 8. A Software Product Line for a family of Instructional Designs

*8.1. A Basic Feature Model*

How to model the mammoth number of instructional designs in a systematic way? Based on patterns [85] and ontologies [86], we present a feature model for modeling a family of instructional designs. Here, we consider standard definitions from SPL literature [96] where a *feature* is a characteristic or end-user-visible behavior of a software system, a *feature model* essentially consists of all the features of a product line and their relationships. A *product* member of a product line is specified by a valid feature selection. Figure §5 shows a generic feature model created using *FeatureIDE* and consists of mandatory features *GoalsPattern*, *ProcessPattern*, *ContentPattern*, *EvaluationPattern*, and optional features *ContextPattern*, *EnvironmentPattern*, which means that any instructional design created from this model must specify these aspects as per the constraints posed in the feature model. For example, the instructional designer has a choice between two ways of specifying goals namely *Bloom* or *ABCD* technique. Figure §6 shows few more details of a feature model for instructional process based on *ProcessPattern* and *ProcessOntology*. However, as specified in Section §8.2, we are interested in feature models with cardinalities, feature attributes and hence we use feature modeling plugin.

*8.2. Feature Attributes*

Feature models primarily specify the features of all product members in a product line primarily from a user perspective. However, if feature models have to be used for (semi-)automatically generating product members or in providing a partial implementation from domain engineering, then feature description alone might not be sufficient and features have to be extended with additional knowledge. For example, to represent *syllables* such as इ, पि in adult literacy, a text in *unicode* should be associated with every feature of that type. Similarly, a goal might have a priority and can be *High*, *Medium*, *Low*. This data can be used by tools during application engineering. However, it was studied that cardinalities and feature attributes make it difficult for verification of valid feature configurations and hence could be useful in only specific domains [108].

In our case, we use cardinalities to impose constraints on the product member and annotate features with data to facilitate further processing by tools. While a feature modeling plugin for *Eclipse* supports feature attributes [106], it is a preliminary prototype developed way back in 2004 and was moved towards the direction of formal verification of features through *Clafer* platform[107]. This need from educational technologies domain requires features to be more powerful and expressive than current notations. This is a future direction beyond this paper and we restrict ourselves to manually annotate features with attributes related to instructional design for our purposes.

*8.3. Product Family Members, Feature Model and Feature Configurations*

Figure §7 shows a brief description of requirements of four different kinds of instructional design specifications for adult literacy. IPCL is the base instructional design for all instructional designs for adult literacy in India. For ID Specification 1, the base ID is provided by IPCL consisting of a set of guidelines for creating primers for all Indian languages based on a core structure, process and content. The essence of IPCL concept is to teach by creating relevant content for learners. Figure §7 shows three concepts namely *Goals*, *Process* and *Content* for different instructional design specifications. The primary goals are Reading, wRiting and aRithmetic at three levels as per the progress of the learners. IPCL describes that an instructional process can be based on synthetic, analytic or eclectic method but suggests use of *eclectic method*. Content is organized as instructional material in the printed primer. There are several primers that are prepared based on this specification and the instructional designer should be able to model them using the product line.

ID Specification 2 family uses instructional design patterns in [85] to describe the Process and Content aspects of the instructional design whereas ID Specification 3 family uses Bloom's revised taxonomy for modeling goals, maps Process and Content patterns to Merrill's principles of instruction. ID Specification 4 does not use the patterns proposed in this paper but uses ABCD technique, Gagne's nine events of instruction for goals and process, and core resources for content. Each of these instructional design specifications can be used to create instructional design editors specific to the family such that instructional designers can create several concrete instructional designs by changing the variabilities in terms of goals, process and content.

In Figure §8, we show a feature model[8] encompassing key features of this product line. This is essentially based on the ontologies for different aspects of instructional design comprising of mandatory features such as *GoalClassification*, *IPCL*, several optional features, selective features and so on. *InstructionaDesignModel* has three choices *MerillModel*, *GagneModel* and *GenericActivty*. Once a teacher chooses *MerillModel*, then *FirstPrinciples* are mandated by default. In case of *GoalPriority*, only one priority out of *High*, *Medium*, *Low* can be chosen. The feature model also mandates that atleast one *Play*, *Act*, *Scene* and *Instruction* are mandatory and can go upto a maximum of 25.

---
[8]A detailed listing of the instructional deign feature model is provided at https://git.io/vdxJG

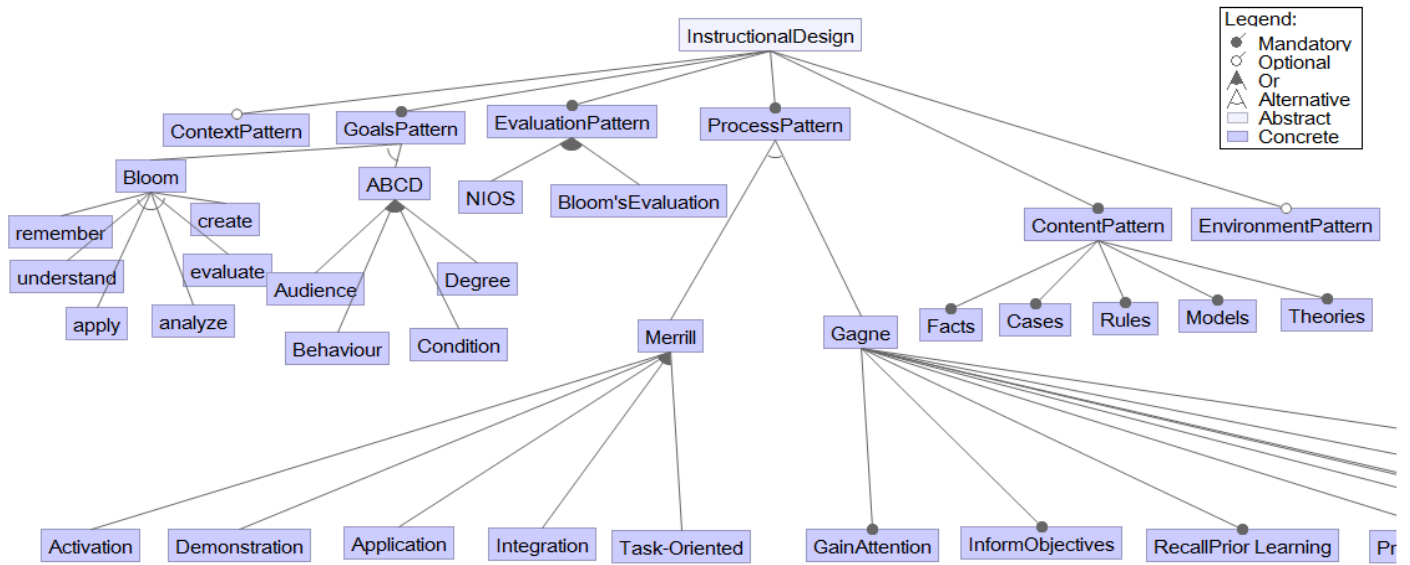

Figure 5: A fragment of instructional design feature model

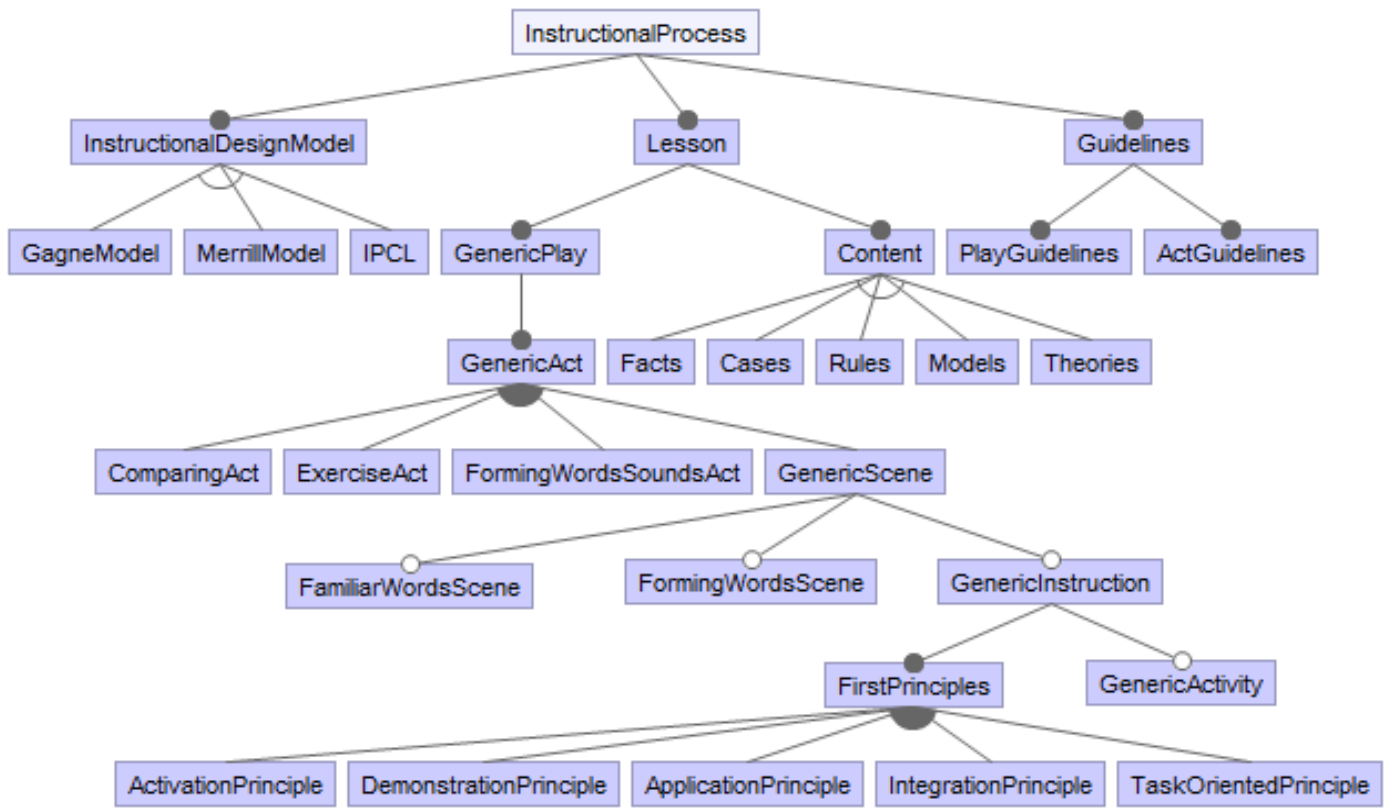

Figure 6: A fragment of instructional process feature model

Figure §9 shows different feature configurations for the different product family members. A *GoalFeature* was configured in three ways as in Figure §9[A,B,C]. Similarly, *ContentFeature* was configured based on specific instructional design requirements. The *ProcessFeature* was con-

| Aspect/ID | ID Specification 1 | ID Specification 2 | ID Specification 3 | ID Specification 4 |
|---|---|---|---|---|
| Base ID | IPCL | IPCL+ *ProcessPattern (pasi)* + *ContentPattern (fcrmt)* | IPCL+ *ProcessPattern (pasi)* + *ContentPattern (fcrmt)* + Merrill's First Principles of Instruction +Bloom's Revised Taxonomy | IPCL + Gagne's Nine levels of learning + ABCD Technique for Goals |
| Goals | *R*ead, w*R*ite and basic a*R*ithmetic (3Rs) | *R*ead, w*R*ite and basic a*R*ithmetic (3Rs) | Organize goals using Bloom's revised taxonomy | Organize goals using ABCD technique |
| Process | Eclectic method for teaching 3*R*s | Organize instructional process as a set of *plays*, *acts*, *scenes*, *instructions* with *instructions* containing actual activities and tasks | Mainly driven by *ProcessPattern (pasi)* but the activities should be based on Merrill's first principles of instruction. | Organize process using Gagne's nine levels of learning |
| Content | Content as per IPCL based primer | Content is organized as *facts*, *cases*, *rules*, *models* and *theories* | Content is organized as *facts*, *cases*, *rules*, *models* and *theories* and mapped to Merrill's first principles | Content is organized as resources |

Figure 7: Custom instructional design specification requirements

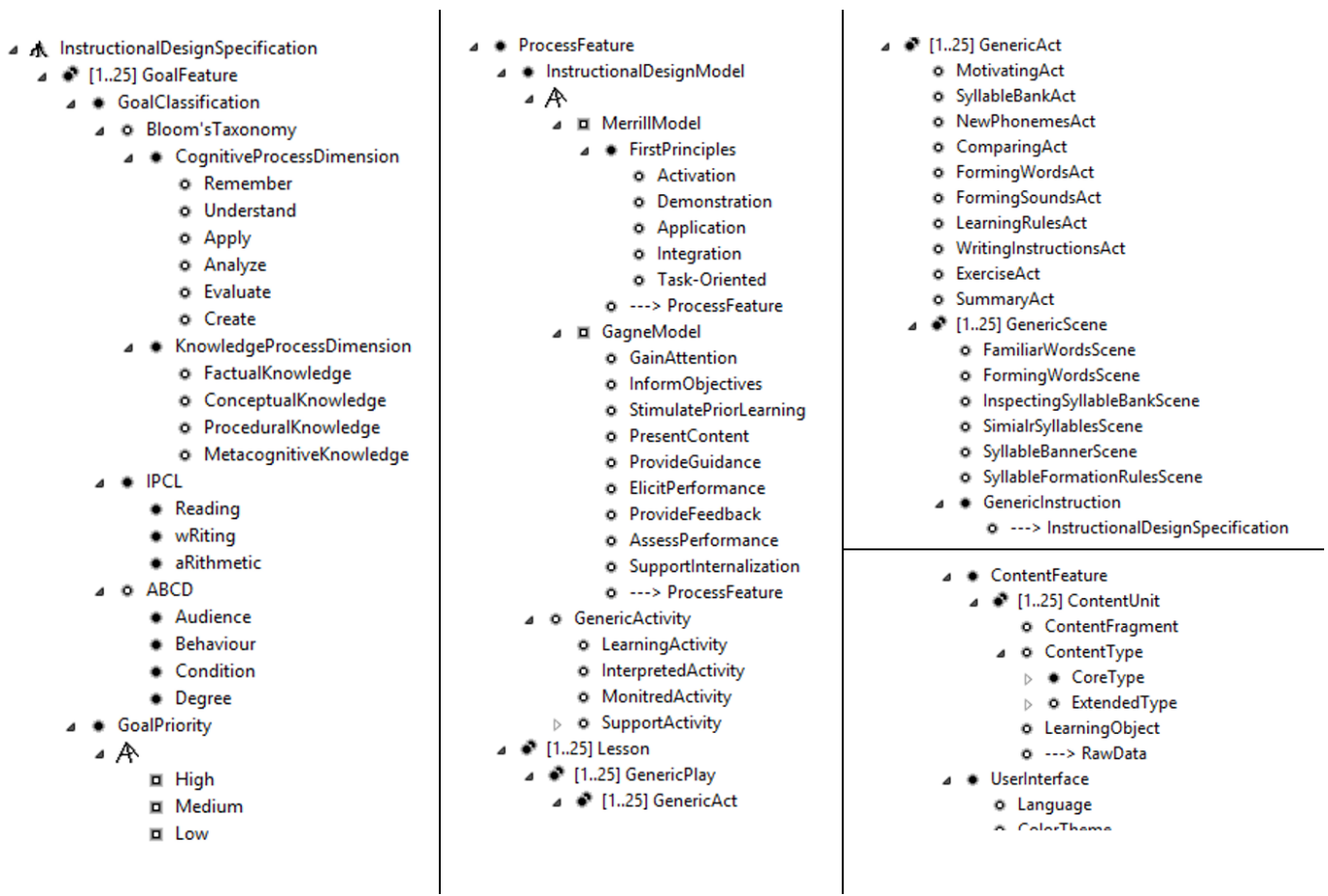

Figure 8: A feature model for instructional design specification

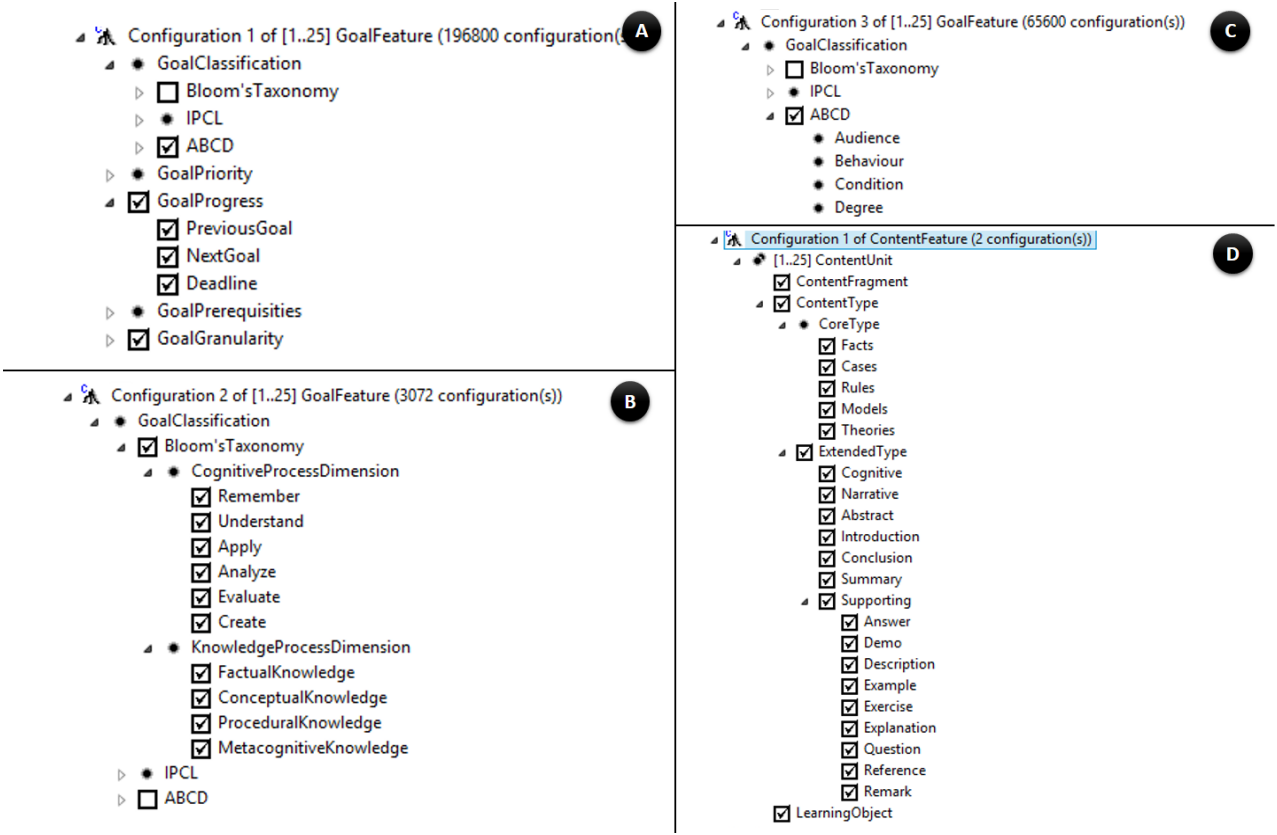

Figure 9: Feature configurations for varied goals and content

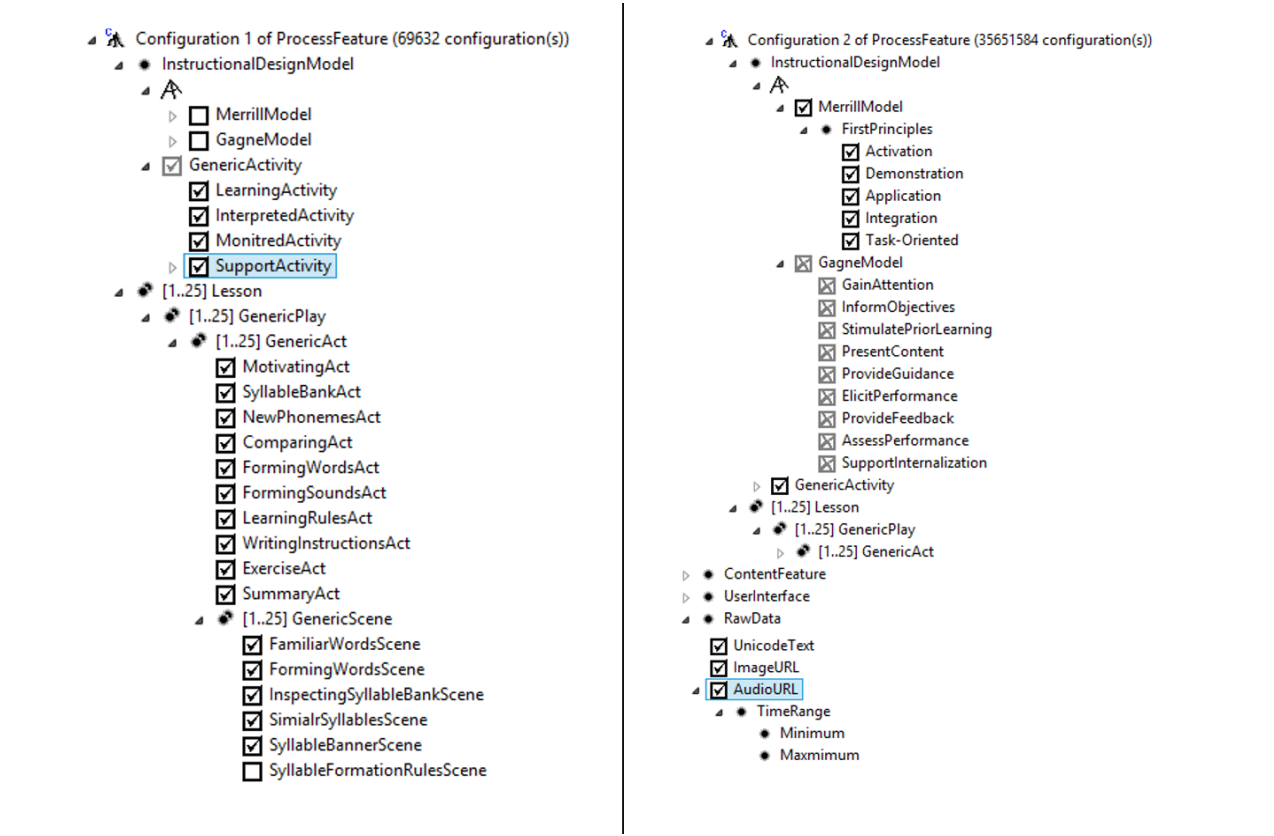

Figure 10: Feature configurations for varied instructional processes

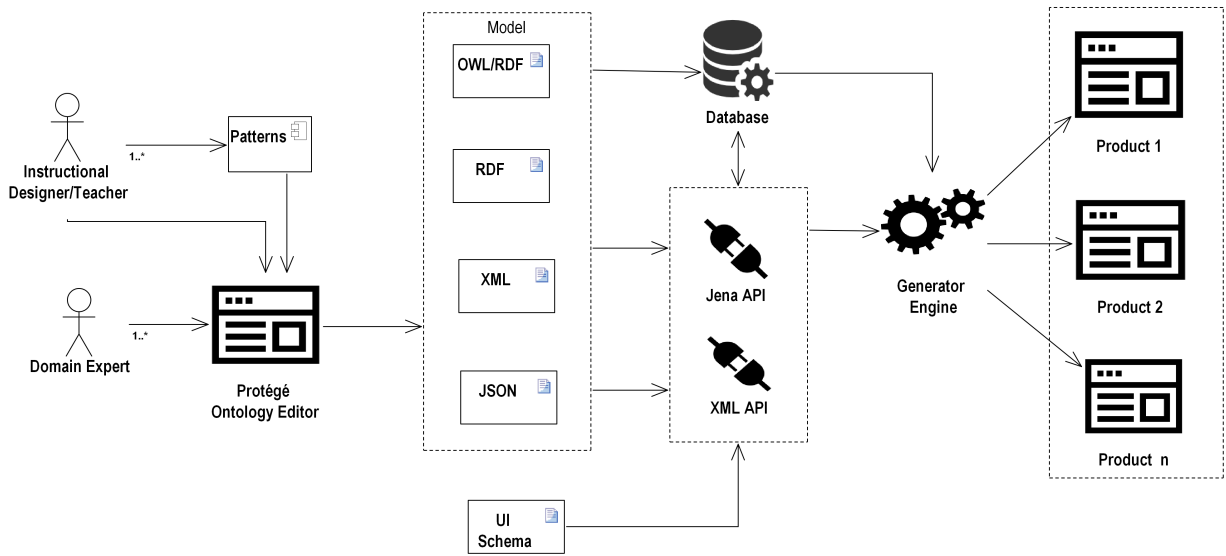

Figure 11: A reference implementation architecture for product lines in this paper

figured in two ways one using the *ProcessPattern* and the other using *MerrillModel* as shown in Figure §10. These possible variations could run into thousands but valid configurations provide different instructional design models with varied goals, processes, content and so on. These custom instructional design specifications are used to create custom authoring tools (editors) for creating instances of the specific instructional design. We first present the reference implementation architecture in the next section followed by a software product line for *e*Learning Systems in Section §9.

### 8.4. A Reference Implementation Architecture

The next step is to take these feature configurations and generate custom instructional design authoring tools (editors) based on specific requirements. One of the key architecture requirements for this product line is that the product family members or web applications should run on limited technical capabilities considering their deployment environment. Internet connectivity cannot be presumed as most of the systems would be in rural villages of India and most of the teachers are either non-technical people or low-computer proficiency teachers. With this constraints, Figure §11 shows a reference implementation architecture for the product lines in this paper. This architecture can be implemented in multiple ways but we discuss our current implementation here. An instructional designer/teacher creates the patterns as document/text and uses that to create an ontology through an ontology editor. We used protégé for creating ontologies in this paper. It can also be the case that an existing ontology be taken. For example, IMS-LD ontology is available in public domain [39] or a comprehensive ontology is available for instructional design teaching learning theories [109]. This ontology can be stored as OWL or RDF file. In addition, we also store ontology as an OWL/XML schema as the current version of platform uses XML for storing knowledge. We also use JSON to store some parts of the OWL or XML for further processing by tools. This data is part of Model in Model-View-Controller pattern. We are currently using Jena API for processing OWL/RDF files and generating a basic web application based on the data in the OWL file. This web application uses the UI schema as input for the generator. We are currently generating two families of applications (product family members) using this architecture. The first set of members are ID editors for selected OWL/XML schema and the generator engine parses the OWL/XML and creates a web application that can be used to create specific instances of instructional design. The other set of applications are *i*Primers or *e*Learning Systems for adult literacy and the generator creates animations based on the specific instructional design described using OWL/XML. This product line is explained in the next section. The current implementation of reference architecture is primary based on files, does not use server but stores all resources in a single package and is implemented mostly using *Javascript, jquery, Nodejs, Jena API*, XML parser, custom animations among others.

The concrete process of creating *ID Editors*[9] is shown in Figure §12. The core input for this process comes in the form of *ID Specifications*, which are created by domain experts. These *ID Specifications* consist of different aspects of instructional design such as *goals*, *process*, *content* based on patterns detailed in [85] and ontologies in [86]. The *ID Editor Product Line* is an engine written in

---

[9]We will use the term ID Editors to mean Instructional Design Editors

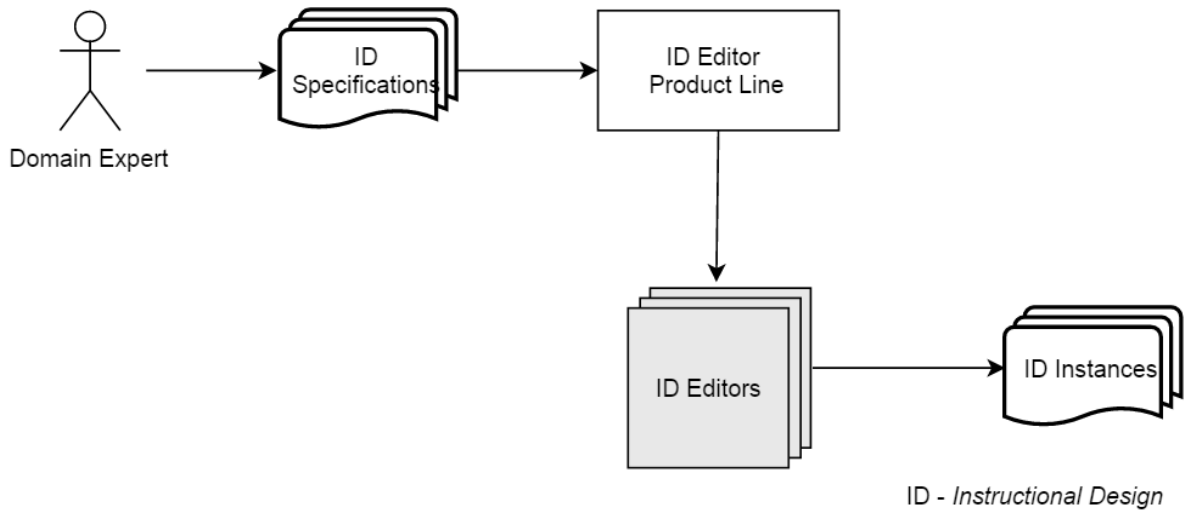

Figure 12: Flow of Instructional Design Product Line

*JavaScript*[10] that parses the *ID Specification* stored in the form of RDF/XML and generates *ID Editors*. This *ID Editor* is a simple form editor consisting of selected aspects of instructional design that are applicable for all instructional design instances based on this concrete specification. This is unlike the current approach of manually creating instructional design editors for every instructional design specification as discussed in Section §7. We have implemented this using multiple technologies such as *Java*[11], *Python*[12]. However, the need has been to create multiple instances of instructional designs which form the basis for several *iPrimers*. We discuss the product line for creating a family of *e*Learning Systems in the next section.

## 9. A Software Product Line for a family of *e*Learning Systems

The primary goal of this product line is to create a family of *e*Learning Systems based on specific instructional designs tailored to the needs of teaching functional literacy for all Indian languages. The 32 State Resource Centers across all states in India are responsible for producing the following primers based on IPCL (first three are mandatory and the rest depend on specific needs) under the aegis of NLMA:

- Basic literacy primer [22+]
- Post literacy primer [22+]
- Life long literacy primers [22+]
- Primers for teaching skills such as tailoring, vocational skills (Jan Shikshan Sansthan (JSS), Life Enrichment Education and so on along with literacy [$n+$, where $n$ is in the order of hundreds]
- Exclusive primers were specifically made for legal literacy, election literacy, agriculture literacy, environment literacy among many others [$n+$, where $n$ is in the order of tens]

An important commonality among these primers is that they teach 3Rs but using varied instructional processes and different themes. Each of these primers are generally available in 22 languages. It is estimated that currently there are atleast 1000 primers available with SRCs in print format. Eventhough the primer is fixed till the next version is developed, officers at different levels (mandal, village, school and teachers) attempt to customize the process, content and adapt it to the local context. For example, a simple way could be to ask the name of learners and find if they know how it looks like? and what are the syllables in it? However, this is not supported in traditional print form, but is a great source of variability for *i*Primers of our product line. How to support immigrants at a given place who want to learn a local language but using their mother tongue as medium of instruction? This leads to another set of variations in the primers with medium of instruction being different for 22 Indian languages?

Technically, we are interested in *i*Primers that are based on field-tested *e*Learning Systems [26]. These applications are based on puppet theater model, where syllables are shown as falling puppets, joining together to form words and so on. The *i*Primers, product members of this family should essentially follow instructional processes, use locally relevant content and present a multimedia application with animations for the learners.

Figure §13 shows the flow of *iPrimer Product Line*. This product line essentially parses the instructional design in-

---

[10]https://github.com/enthusiastic2learn/ID-Editor
[11]https://github.com/enthusiastic2learn/IDPlatform
[12]An implementation named *Semantic Web Forms* was done by undergrad students as part of Software Engineering course at IIIT-Sri City, India and is available at https://github.com/chrizandr/semantic_web/

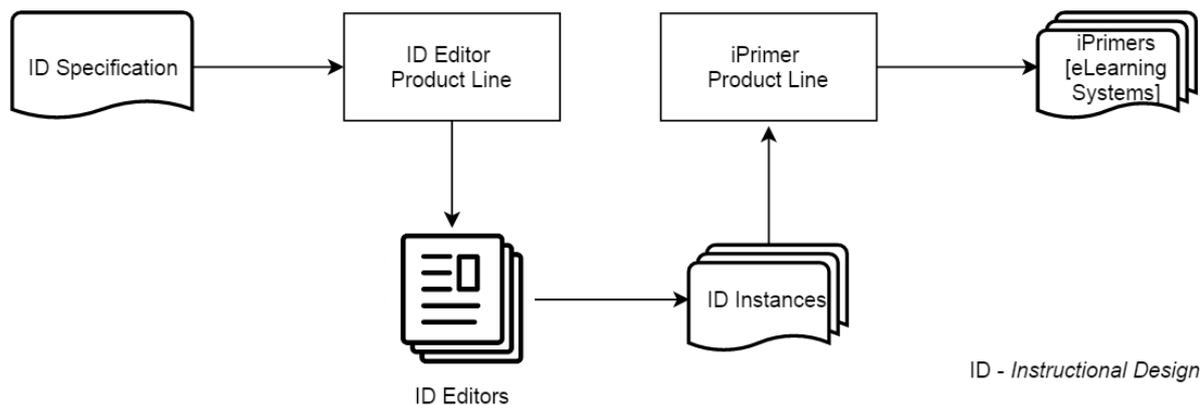

Figure 13: Flow of *i*Primer Product Line

stances to generate *i*Primers. In the case of adult literacy in India, the *iPrimer Product Line* is based on a single *ID Specification* driven by IPCL. This *ID Editor* is used to create several instances of the instructional design specification for varied processes, content and visual and audio elements. These instances are parsed by *iPrimer Product Line* to eventually create *i*Primers for multiple languages and primers. Every instructional design instance leads to a varied *i*Primer of the product line. The *iPrimer Product Line* has to be customized if the base *ID Specification* is changed to other than pre-defined *ID Specification* as the RDF/XML parser has to be re-written and it would take about a person-week to re-write the parser for ID specifications beyond adult literacy. Section §10 presents the results of cost savings of *ID Editor Product Line* and *iPrimer Product Line*. We used the *iPrimer Product Line* to generate several *i*Primers and discuss *i*Primers for *Hindi* and *Telugu* Language in this section.

Figure §14 shows a fragment of primer of *Hindi* language. This primer has around 180 pages with 24 lessons and each lesson teaching 3Rs. This primer is available in both print as well as digitized format (pdf). This digitized form is used as an input to a custom instructional design editor for creating a custom instructional design instance as shown on the right hand side. This OWL/XML file[13] contains all the information related to a specific instructional design and serves as the base for creating variations based on this instructional design. Figure §15 shows how some variations can be created using the *iPrimer Product Line*. The *iPrimer Product Line* primarily reads the OWL/XML file for instructional process consisting of activities, their order, and content that has to be used in the process and generates animations accordingly. Everything that is shown in Figure §15 can be varied as per the feature model configurations discussed in earlier sections. This allows to rapidly customize the *i*Primers and create new ones by changing processes and content. Figure §16 shows how an *i*Primer has been generated for *Telugu* language based on a specific instructional design instance[14]. Here, the processes, content, user interface that are relevant for that specific instructional design have been generated. Figure §17 shows some variations that are possible for *Telugu* language. The core idea here is to be able to generate as many *i*Primers as possible with minimum effort by applying the idea of software product lines. We have observed that this product line can be configured easily to create *i*Primers but one major obstacle is with respect to sound, which has to be created manually in the current version. However, we are thinking of using teachers'/learners' voice to record instructions and content at a personalized level as part of our future work.

## 10. Experimental Results

In this section, we discuss the experimental results of using our approach and technologies for (semi-)automatic creation of *ID Editors* and *i*Primers. We wish to reiterate that the primary goal of this paper is to facilitate customization of educational technologies for scale and variety and demonstrate it in the context of adult literacy in India. One of the core claims of software product lines is that product lines facilitate creation of product variants at reduced cost [60]. The literature has a number of measures to calculate the cost and return on investment on software product lines [110]. In this paper, we consider the commonly used model of Structured Intuitive Model for Product Line Economics (SIMPLE) to measure the effectiveness of product lines [60]. The SIMPLE model describes seven scenarios for creation of SPLs that may typically occur in an organization. The generic scenario is concerned with creation of SPLs and stand alone products from existing products and resources. Specifically, the SPLs in this paper fall into the category of *Scenario 2*, where the organization plans to develop a set of products as a product line based on common core assets. The SIMPLE model consists of four cost components to calculate the total cost of SPLs [110].

---

[13]A detailed listing of this instructional design instance is available at https://git.io/vdxJV

[14]Available at https://git.io/vdxJK

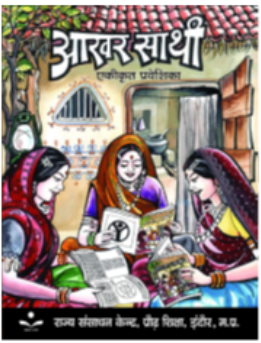

Figure 14: Primer and custom instructional design instance [XML from OWL] for *Hindi* language

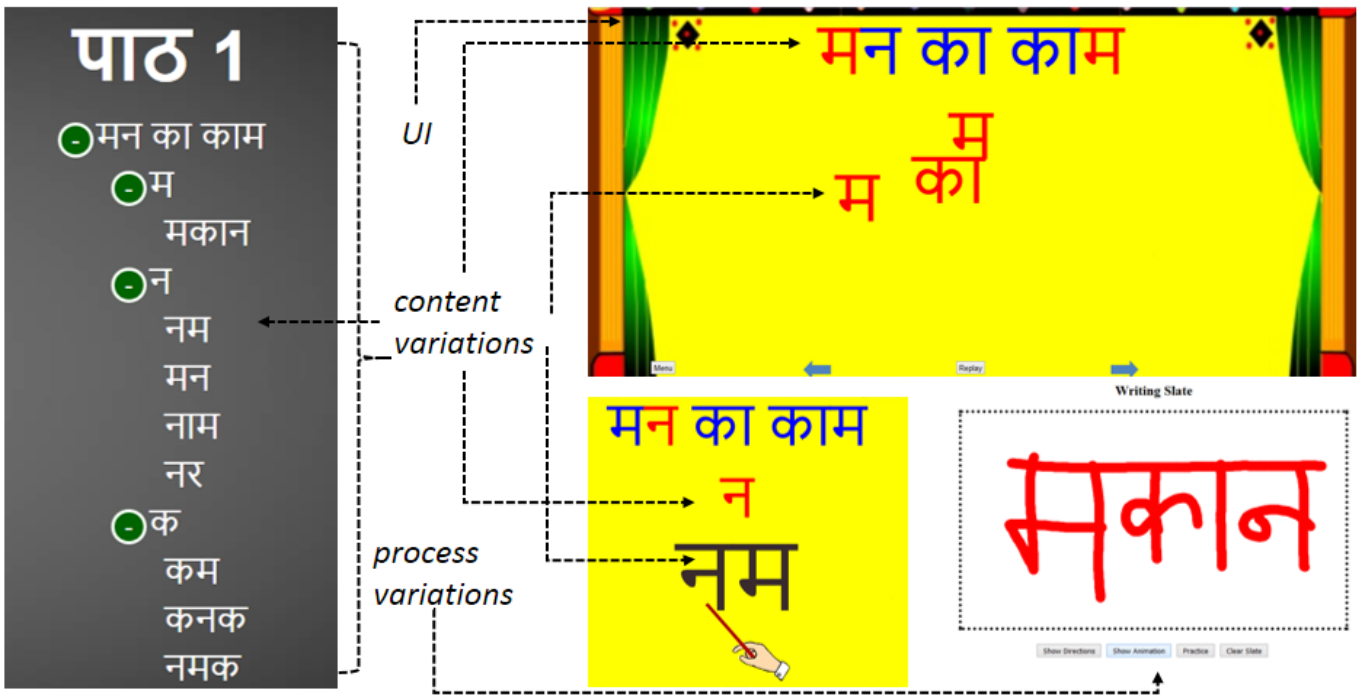

Figure 15: *i*Primer for *Hindi* language - generated from instructional design instance

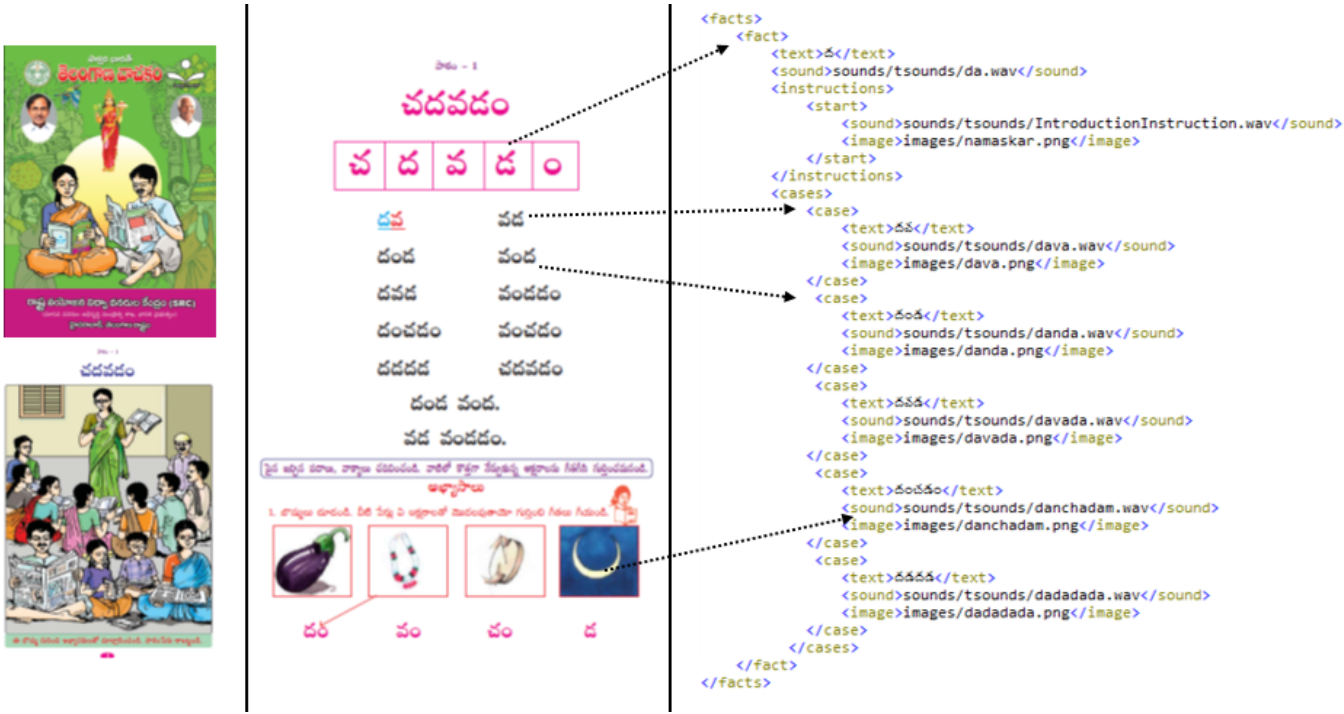

Figure 16: Primer and custom instructional design instance [XML from OWL] for *Telugu* language

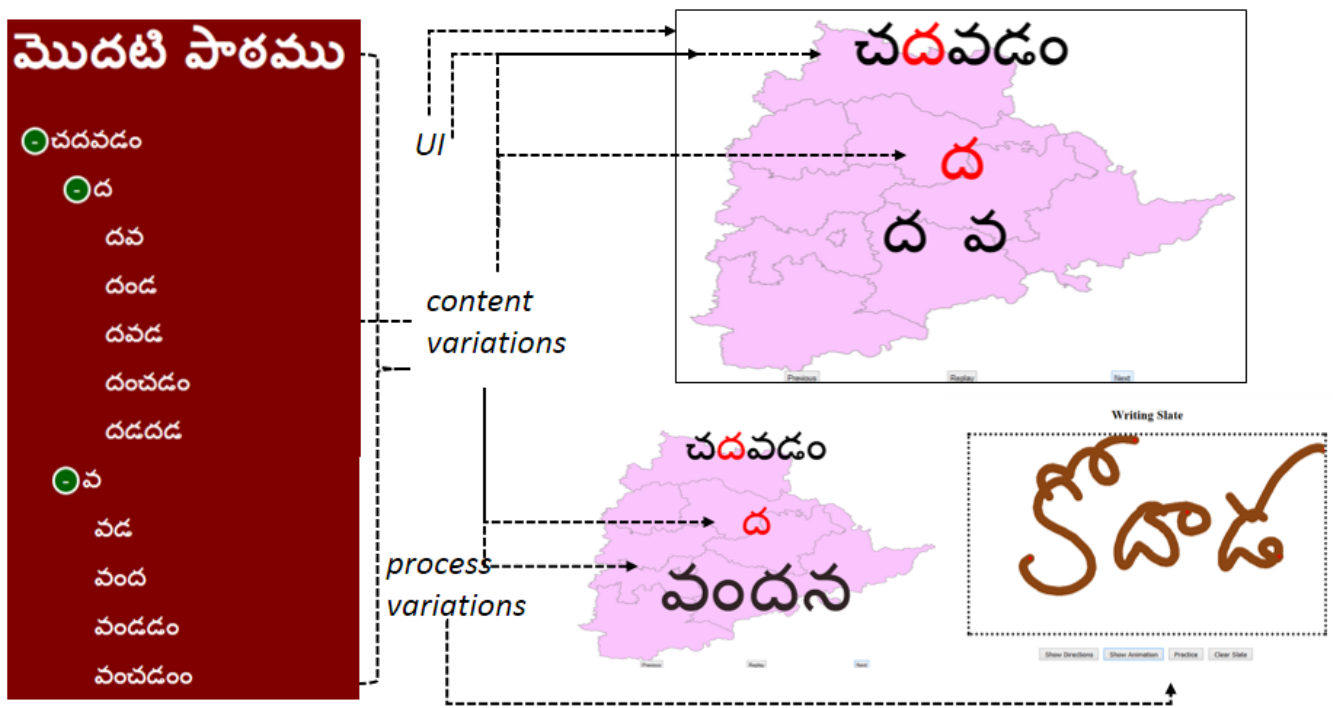

Figure 17: *i*Primer for *Telugu* language - generated from instructional design instance

- $C_{org}$ - The cost to an organization for adopting product line approach instead of single system development. In this paper, the product lines are developed by researchers and hence no direct organization costs. However, in the long run, the organization that develops software for all *i*Primers should incur costs for transition to product line approach.

- $C_{cab}$ - The cost to develop core assets that are reusable across the product line. This cost includes the patterns discovered, ontologies created along with traditional SPL activities.

- $C_{unique}$ - The cost to develop unique features of the product beyond the product line. This generally involves manual effort to customize the generated product from the product line.

- $C_{reuse}$ - The cost to reuse core assets, adapt them for the needs of developing new products in the product line.

The costs of developing a software product line for $n$ distinct products can be calculated as follows [111][112]:

Cost of building a product line
$$C_{SPL} = C_{org}() + C_{cab}() + \sum_{i=1}^{n}(C_{unique}(product_i) + C_{reuse}(product_i))$$

Cost of building $n$ stand-alone products
$$C_{stand-alone} = \sum_{i=1}^{n} C_{product}(product_i)$$

where $C_{product}$ is the cost of developing an individual product.

The savings of software product lines can be estimated as:
Savings of product lines = $C_{stand-alone}$ - $C_{SPL}$

*Tata Consultancy Services*, an Indian software services organization has been involved with development of *e*Learning Systems for adult literacy in India for more than 15 years [26]. We use data from our earlier experience of developing *e*Learning Systems [20] and TCS' statistics on developing *e*Learning Systems for 9 Indian Languages[26] as the initial base for calculating cost savings of *iPrimer Product Line*. The effort for creating an *e*Learning System was around 5 to 6 person years and in our earlier work, we have applied software reuse techniques and reduced the effort for creating *e*Learning Systems to 5 to 6 person months [20]. Each existing *i*Primer approximately consists of 20,000 visual elements; 2,500 sound elements with 500 words based on a physical primer for a language. These elements are organized in the form of approximately 24 lessons constituting an *e*Learning System for teaching

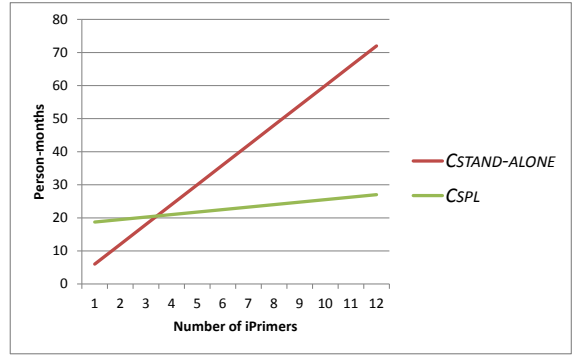

Figure 18: Cost Savings of *iPrimer Product Line*

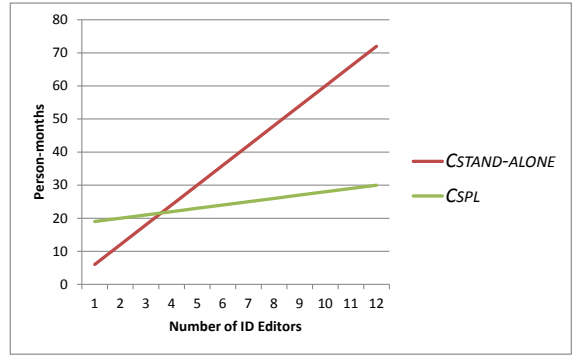

Figure 19: Cost Savings of *ID Editor Product Line*

3Rs. The *iPrimer Product Line* essentially generates these 24 lessons as shown in Figure §15 and Figure §17 for *Hindi* and *Telugu* languages with manual inputs for words and sounds. Based on this existing data, we evaluate the cost savings of *iPrimer Product Line* as follows:

Here, we present the costs for building 9 products i.e., *i*Primers:

Cost of building a product line
$C_{SPL}$ = 6 person-months + 12 person-months + 9 * (2 person-weeks + 1 person-week)
$C_{SPL}$ = 25 person-months

Cost of building $n$ stand-alone products
$C_{stand-alone}$ = 9 * 6 person-months
$C_{stand-alone}$ = 54 person-months

where $C_{product}$, the cost of developing an individual product is 6 person-months.

The savings of software product lines can be estimated as:

Savings of product lines = 54 person-months - 25 person-months i.e., 29 person-months horizontal axis shows the number of *i*Primers and the vertical axis shows the number of person-months required to develop the *i*Primers. The graph shows that the break-even for the initial investment in terms of core asset base is for 3 or 4 *i*Primers after which as the number of *i*Primers to be developed increases, the cost required for developing them in a stand-alone fashion increases rapidly whereas it is steady in the case of SPL. The *iPrimer Product Line* hosted at http://rice.iiit.ac.in was used by a low-computer proficiency teacher at State Resource Center, Telangana, India to create 10 lessons of *i*Primer based a newly released physical primer. This *i*Primer was packaged as an android app using *Apache Cordova*[15] and is hosted online at Google Play Store Store[16]. The low-computer proficiency teacher was able to create these lessons in about a day but without audio and the instructional design instance created using the *iPrimer Product Line* is available on *Github* [17]. The primer was also listed on Government of Telangana websites[18]. In addition, a workshop was conducted in November 2016 for 24 preraks of adult literacy on the use of *iPrimer Product Line*. Figure §20 shows a glimpse of the session where teachers used *iPrimer Product Line* to create partial lessons based on dynamic words given by the audience.

Table §2 shows the individual cost components for *iPrimer Product Line* and Figure §18 shows the cost of creating *i*Primers with and without our approach. The

Table 2: Cost components of *iPrimer Product Line*

| Cost Component | Cost (Person-months) | Description |
|---|---|---|
| $C_{org}()$ | 6 person-months | In case of *iPrimer Product Line*, we do not have a single organization but we have developed the product line as part of this paper essentially meaning no direct cost for an organization to adopt the product line approach. However, based on our experience and collaboration with TCS, we consider a time of 6 person-months as an organizational cost. |
| $C_{cab}()$ | 12 person-months | Core assets in the case of *iPrimer Product Line* are ontologies of instructional design that were developed based on patterns, which are represented in RDF/OWL format, *JavaScript* files, a parser that reads configuration files as an XML and generates instances, UI components such as animation generator and so on. We have spent around 12 person-months to create this core asset base which is part of the reusable infrastructure of this product line. |
| $C_{unique}()$ | 2 person-weeks | The unique parts of the *i*Primers are primarily *process* steps and *content* in terms of words, syllables, which have to be extracted from a soft copy of the primer or to be entered manually. In addition, the software has to be adapted to handle special syllables or words that are specific to the particular language. The cost to create sound files for new words is a major source of manual effort as text-to-speech tools for Indian Languages are not yet acceptable for purposes of literacy teaching. |
| $C_{reuse}()$ | 1 person-week | The cost to modify existing resources i.e., instructional design instance with data or raw XML aspects for user interface elements pertaining to a specific *i*Primer. |

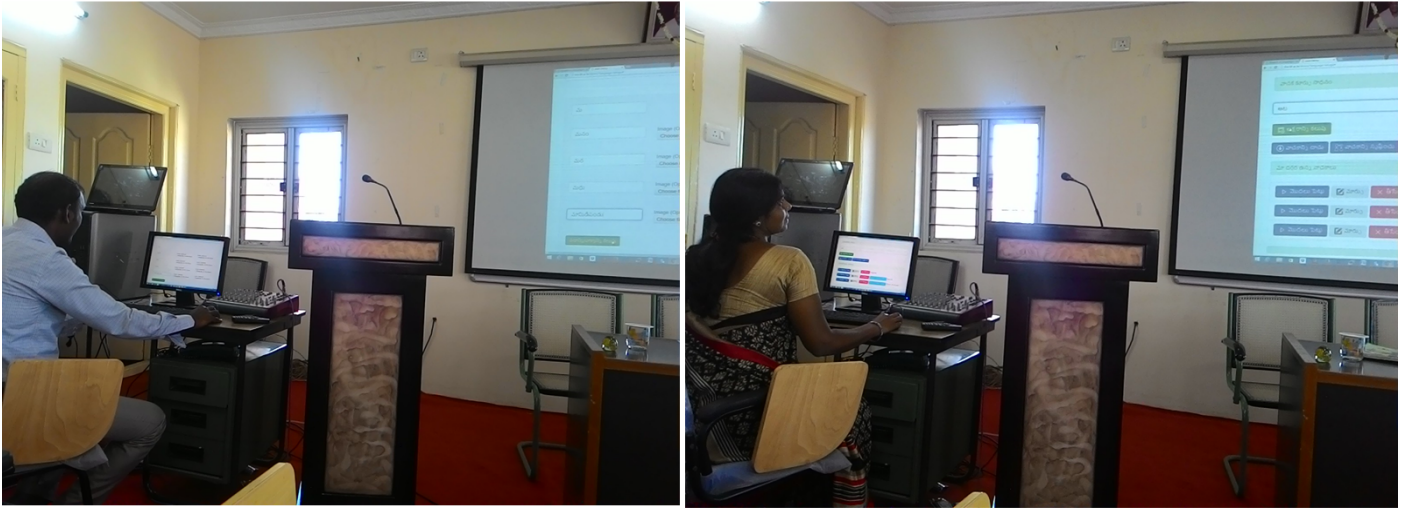

Figure 20: Teachers using *iPrimer Product Line* to create *i*Primers

---
[15]https://cordova.apache.org
[16]https://goo.gl/6LBBtq

We have also populated the cost of developing *ID Editors* with and without SPL in Figure §19. Here, the cost of manual effort for customizing the generated *ID Editor* is one person-month instead of three person-weeks as in *iPrimer Product Line*.

## 11. Conclusions

In this paper, we aimed at creating an approach for design and customization of educational technologies to address *scale* and *variety* in education. Specifically, we addressed the need to support creation of *e*Learning Systems for flexible instructional designs and multiple Indian Languages in the context of adult literacy in India. We explained the development of software product lines for a family of (i) instructional designs (ii) *e*Learning Systems and discussed how these are connected to each other through a reference architecture. We demonstrated our approach by creating an *ID Editor Product Line* and *iPrimer Product Line* and further semi-automatically generated *e*Learning Systems for adult literacy case study. The work presented in this article is one of the first attempts of large scale application of software engineering approaches for educational technologies and can lead to a significant line of research in the area of software reuse and software product lines.

## 12. Future Work - Software Product Lines for Personalized Learning

Even though the software product line approach outlined in this paper is a natural way to address scale and variety of educational technologies, personalized learning is a grand challenge for computing requiring further research from software product lines community.

- Using current feature modeling notations, features can only be selected for product configuration but the need in educational technologies is to have features that have knowledge associated with them for different aspects of instructional design such as goals, process steps and content, which is not possible with current notations. For example, expressing goals using Bloom's taxonomy or ABCD technique could be a feature but specifying an exact learning goal requires more than just features.

- Design of light-weight approaches for SPL for educational technologies domain is a definite need as instructional design itself is a complex activity.

- Educational technologies domain presents the need for a family of product lines catering to the needs for variety at multiple levels.

- The socio-technical nature of education domain motivates the need for SPLs that are spread across domains such as learning methodologies, software engineering and human-computer interaction.

- Design of SPLs that span across different organizations from different domains.

- Facilitating assembly of educational technologies from open educational resources and further customizing them for personalized learning requires research in every aspect of software product lines from scoping to all aspects of domain and application engineering.

- In addition, lean and globally distributed software product lines could be two potential research directions for addressing the challenges of designing educational technologies for personalized learning.

From application perspective, software product lines can be developed for multiple domains in education such as schools, skills and different forms of engineering and medical education.

## Acknowledgements

We would like to thank TCS for providing us with initial inputs for this work, NLM for taking our work forward to create national impact, Government of Telangana for being one of the first adoptors of our technologies and all funding agencies for supporting several international research travels during this research.

---

[17]https://git.io/vdxkd
[18]http://tslma.nic.in/ and State Resource Center, Government of Telangana at http://srctelangana.com/